\documentclass[letter]{aa}  

\usepackage{graphicx}
\usepackage{txfonts}
\usepackage[colorlinks=true,allcolors=blue]{hyperref}
\usepackage{subcaption}
\def\HII{H\,{\sc{ii}}}

\def\herschel{$\it{Herschel}$\,}

\def\getsources{$\it{getsources}$}
\def\getimages{$\it{getimages}$}

\begin{document}

   \title{ALMA observations of RCW~120}

   \subtitle{Fragmentation at 0.01~pc scale\thanks{The \getsources\,catalog together with the configuration file are available at the CDS via anonymous ftp to *** or via ***}}

   \author{ M. Figueira\inst{1,2} 
            \and L. Bronfman\inst{2}  
            \and A. Zavagno\inst{1}
            \and F. Louvet\inst{2}  
            \and N. Lo\inst{2}
            \and R. Finger\inst{2}
            \and J. Rodón\inst{3}           
}

 %\offprints{M. Figueira, miguel.figueira@lam.fr }

   \institute{Aix Marseille Univ, CNRS, LAM, Laboratoire d'Astrophysique de Marseille, Marseille, France
   \and Departamento de Astronomía, Universidad de Chile, Casilla 36-D, Santiago, Chile
   \and Onsala Space Observatory, Sweden
     }

   \date{Received 3 March 2018 ; accepted ***}

% \abstract{}{}{}{}{} 
% 5 {} token are mandatory
 
  \abstract
   {Little is known about how high-mass stars form. Around 30\% of the young high-mass stars in the Galaxy are observed at the edges of ionized (\HII) regions. Therefore these are places of choice to study the earliest stages of high-mass star formation, especially towards the most massive condensations. High-spatial resolution observations in the millimeter range might reveal how these stars form and how they assemble their mass.}
   {We want to study the fragmentation process down to the 0.01~pc scale in the most massive condensation (1700~$M_{\odot}$) observed at the south-western edge of the \HII\,region RCW~120 where the most massive \herschel cores ($\sim$124~$M_{\odot}$ in average) could form high-mass stars.}
   {Using ALMA 3~mm continuum observations towards the densest and most massive millimetric condensation (Condensation 1) of RCW~120, we used the \getimages\,and \getsources\,algorithms to extract the sources detected with ALMA and obtained their physical parameters. The fragmentation of the \herschel cores is discussed through their Jeans mass to understand the properties of the future stars.}
   {We extracted 18 fragments from the ALMA continuum observation at 3~mm towards 8 cores detected with $\it{Herschel}$, whose mass and deconvolved size range from 2~$M_{\odot}$ to 32~$M_{\odot}$ and from 1.6~mpc to 28.8~mpc, respectively. The low degree of fragmentation observed, regarding to the thermal Jeans fragmentation, suggests that the observed fragmentation is inconsistent with ideal gravitational fragmentation and other ingredients such as turbulence or magnetic fields should be added in order to explain it.
    Finally, the range of fragments' mass indicates that the densest condensation of RCW~120 is a favourable place for the formation of high-mass stars with the presence of a probable UC\HII\, region associated with the 27~$M_{\odot}$ Fragment 1 of Core 2.}
   {}

   \maketitle
%
%-------------------------------------------------------------------

\section{Introduction}

Despite decades of simulations and observational studies, the early-stages of high-mass star formation are still poorly understood. Because high-mass stars shape the native cloud in which they form through feedback processes such as momentum injection through winds, heating and photoionization via radiation, and violent supernova explosions, a deeper understanding of their whole evolution scenario is important. High-mass stars are known to form within gas condensation called clump ($\sim$ 1~pc scale) and cores ($\sim$ 0.1~pc scale). They also form ionized (H\,{\sc{ii}}) regions whose expansion collects the surrounding molecular material which can have positive (triggering mechanisms, \citealt{elm77,kes03}) or negative impact \citep{dal05,luc17} on the local star formation. Despite the fact that about 30\% of the high-mass star formation in the Galaxy is observed at the edges of these \HII\, regions \citep{deh10,ken12,ken16,pal17}, the way these stars form and the efficiency of the possible triggering mechanisms are still debated \citep{dal15,pal17}. Several observations of cores at a spatial resolution of 0.01~pc \citep{mot98,bon10,pal15,oha18,pal18} were performed to study the properties of star formation but the lack of observations around \HII\,regions prevent the study of their impact on a new generation of (high-mass) stars. Recent ALMA observations towards W43-MM1 at 2400~AU resolution \citep{mot18} suggests that the CMF is top-heavy and not self-similar to the IMF contrary to low-mass star-forming regions such as $\rho$-Ophiuchi or the Aquila complex \citep{mot98,kon15}. Moreover, studies have shown that turbulence \citep{pad01} and magnetic fields \citep{hen08} are key ingredients that allow the structures to support the gravitational collapse above the thermal Jeans mass, allowing the formation of massive fragments ending with one or binary high-mass stars. The Galactic \HII\,region RCW~120 \citep{zav07,deh09} represents a text-book example to further study the properties of star formation observed at its edges, where the most massive \herschel cores extracted and studied by \citet{fig17} were observed with ALMA. In Section~\ref{sect:rcw120}, we present the RCW~120 region and the Condensation 1, Section~\ref{sect:observations} describes the ALMA observations used in this work with the data reduction process, Section~\ref{sect:results} presents the analysis of the fragments and their properties, Section~\ref{subsect:fragmentation}, \ref{subsect:massivestars_rcw120} discusses the fragmentation of the \herschel cores relative to the Jeans mass and the probability that high-mass stars emerge towards the PDR. Finally, Section~\ref{sect:conclusions} presents the conclusions of this work.

\section{The Galactic \HII\,region RCW~120}\label{sect:rcw120}

\begin{figure}
\begin{subfigure}{0.5\textwidth}
\centering
  \includegraphics[width=\linewidth]{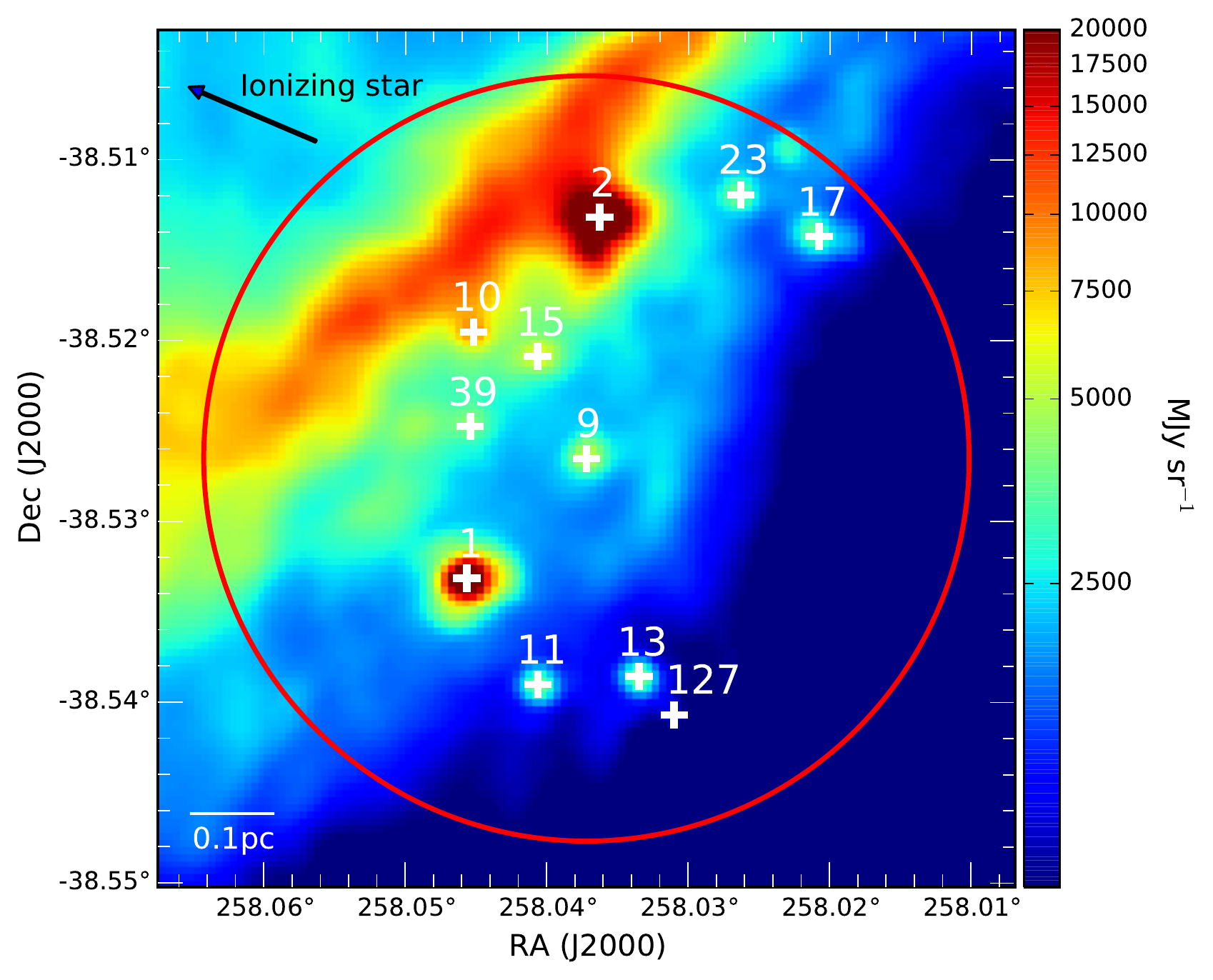}
  \label{fig:sub2}
\end{subfigure}
\begin{subfigure}{0.5\textwidth}
\centering
  \includegraphics[width=\linewidth]{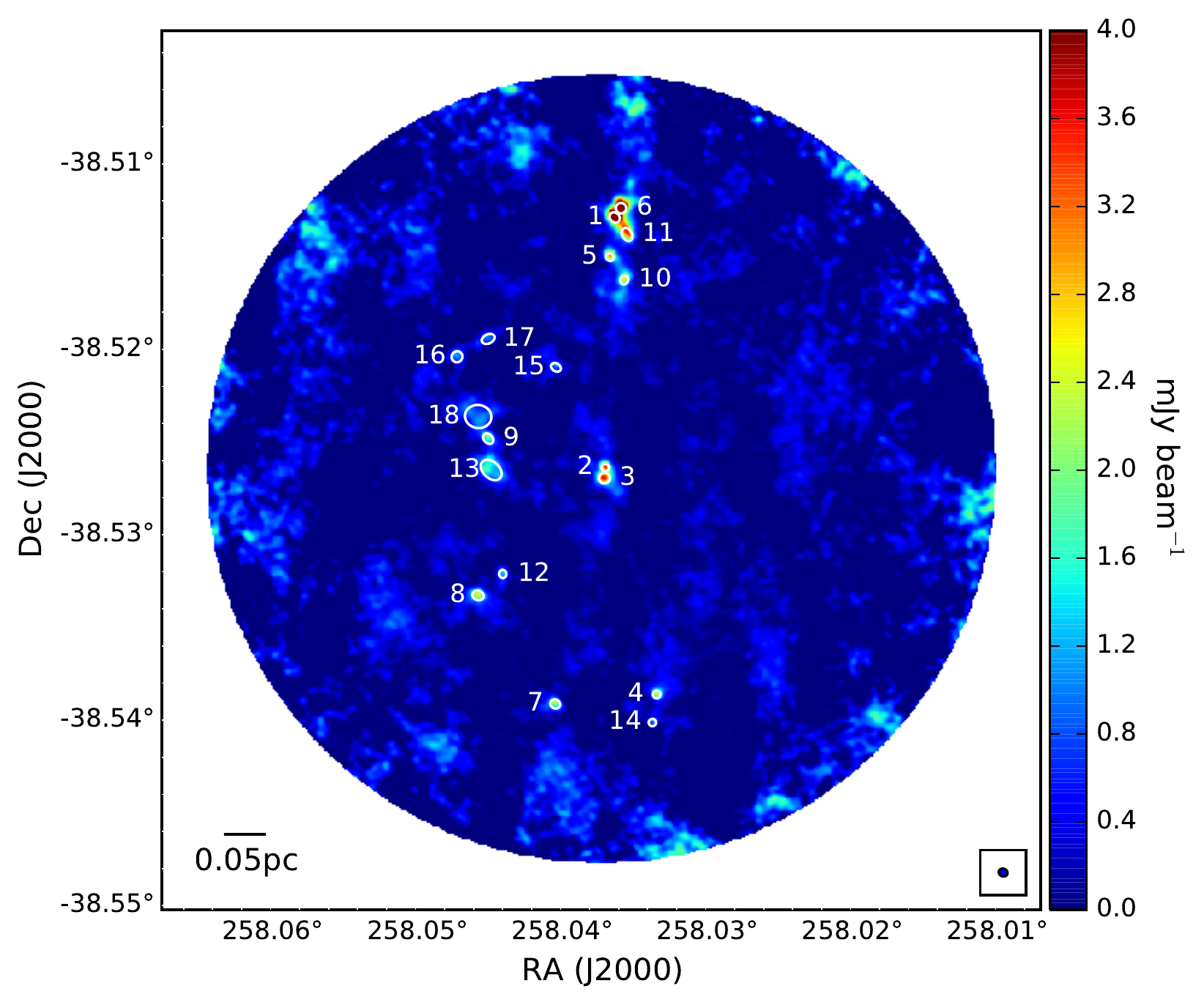}
  \label{fig:sub3}
\end{subfigure}
\caption{Top: \herschel\,70~$\mu$m image of Condensation 1 with the sources detected with the \getimages-\getsources\,algorithms from \citet{fig17}, the FoV of the ALMA observation presented here (red circle) and the direction where the ionizing star is located (black arrow). Bottom: 3~mm ALMA image of the fragments observed towards Condensation 1.}
\label{fig:rcw120_alma}
\end{figure}

RCW~120 \citep{rod60} is a Galactic \HII\,region ionized by a single O8V star \citep{mar10} and located 0\fdg5 above the Galactic Plane at a distance of 1.3~kpc. Studied with $\it{Spitzer}$, APEX-LABOCA and SEST-SIMBA \citep{zav07,deh09}, the latest study was made by \citet{fig17} using the \herschel observations of the OB Young Stellar objects (HOBYS) guaranteed time key program (\citealt{mot10}). A sample of 35 cores were extracted using the \getsources\,algorithm \citep{men12,men13} from which the most massive and youngest ones are located towards the densest condensation (Condensation 1: 1700~$M_{\odot}$ in 0.7~pc $\times$ 0.5~pc, see Fig.~\ref{fig:rcw120_alma} top). Using the ALMA 12~m antennas, we observed the massive sources seen towards this condensation in order to understand how the \herschel\,cores are fragmented at 0.01~pc. In particular, we aim to understand if massive stars can be formed at the edges of this \HII\,region. 

\section{ALMA observations and data reduction}\label{sect:observations}

The observations of the Condensation 1 were performed during Cycle 4 using 38 of the 40 12~m ALMA antennas in nominal configuration C40-3 with baselines ranging from 15~m to 459~m. We used a continuum bandwith of 2.227~GHz in Band 3 divided into 3$\times$117.19~MHz (122~kHz$-$0.4~km~s$^{-1}$ resolution) + 1$\times$1875~MHz bandwidth. The spectral bands were centered on 93.17, 91.98, 104.02 and 102.5~GHz allowing the observations of N$_2$H$^+$, CH$_3$CN, SO$_2$ transition lines and 3~mm continuum emission, respectively. The primary beam at the sky representative frequency of 104.03~GHz is 56\arcsec. The total observing time was 27.8~minutes with a system temperature between 46.7~K and 69.9~K and an average precipitable water vapor of 1.6~mm. For this work, we used the products delivered by ALMA. The imaging was performed with the CLEAN algorithm of CASA using a Briggs weighting with a robust parameter of 0.5 and a primary beam limit of 20\%. The continuum emission was subtracted from all the spectral data cubes and we ended-up with a synthesized beam of 1\farcs7$\times$1\farcs5 (0.01~pc at a distance of 1.3~kpc) and a rms noise level of 0.16~mJy~beam$^{-1}$ for the aggregate continuum (4 spectral windows) and 4.6~mJy~beam$^{-1}$ for the SO$_2$ and CH$_3$CN molecular line transitions. After an inspection of the subtracted continuum and the residual maps for each spectral window, the images made by the pipeline appear to be of good quality. The largest recoverable angular scale for these observations is 14\arcsec. In this letter, we present the high-resolution 3~mm continuum observation towards Condensation 1 together with SO$_2$ and CH$_3$CN spectroscopic observations towards one particular fragment. The other data will be presented in a forthcoming paper (Figueira et al., in prep.).

   \begin{table*}    
\centering          
\begin{tabular}{|ccccrrcccr|}     % 7 columns 
\hline
Id & $\alpha$ & $\delta$ & T$_{\rm{env}}$ & M$_{\rm{env}}$ & L$_{\rm{bol}}$  & n$_{H_2}$\tablefootmark{a} & M$_{\rm{Jeans}}$ & N$_{\rm{frag}}$ & M$_{\rm{frag}}$  \\
  & \multicolumn{2}{c}{J2000 ($^{\circ}$)} & (K)  & ($M_{\odot}$) & ($L_{\odot}$) & (cm$^{-3}$) & ($M_{\odot}$) & & ($M_{\odot}$) \\
\hline
1	&	258.04577	&	-38.53338	&	17.0	$\pm$	0.2	&	85	$\pm$	6	&	234	$\pm$	28	&	(3.0$\pm$0.2)	$\times$10$^{5}$	&	0.8	&	2	&	10.6	$\pm$	0.7	\\
2	&	258.03624	&	-38.51317	&	16.9	$\pm$	0.2	&	376	$\pm$	21	&	856	$\pm$	93	&	(1.3$\pm$0.1)	$\times$10$^{6}$	&	0.4	&	5	&	73	$\pm$	3.6	\\
9	&	258.03749	&	-38.52663	&	13.1	$\pm$	0.2	&	97	$\pm$	14	&	49	$\pm$	12	&	(3.4$\pm$0.5)	$\times$10$^{5}$	&	0.5	&	2	&	25.8	$\pm$	1.6	\\
10	&	258.04524	&	-38.51956	&	11.1	$\pm$	0.4	&	252	$\pm$	41	&	46	$\pm$	17	&	(8.7$\pm$1.4)	$\times$10$^{5}$	&	0.3	&	2	&	15.5	$\pm$	1.4	\\
11	&	258.04073	&	-38.53926	&	14.2	$\pm$	0.4	&	31	$\pm$	9	&	24	$\pm$	11	&	(1.1$\pm$0.3)	$\times$10$^{5}$	&	1.1	&	1	&	7.4	$\pm$	0.5	\\
13	&	258.03352	&	-38.53886	&	16.3	$\pm$	0.8	&	8	$\pm$	3	&	23	$\pm$	9	&	(2.8$\pm$1.0)	$\times$10$^{4}$	&	2.5	&	2	&	6.8	$\pm$	0.8	\\
15\tablefootmark{b}	&	258.04084	&	-38.52110	&	12.8	$\pm$	0.5	&	81	$\pm$	15	&	38	$\pm$	15	&	(2.8$\pm$0.5)	$\times$10$^{5}$	&	0.5	&	1	&	3.5	$\pm$	0.6	\\
39	&	258.04560	&	-38.52532	&	12.8	$\pm$	0.3	&	97	$\pm$	17	&	42	$\pm$	13	&	(3.4$\pm$0.5)	$\times$10$^{5}$	&	0.5	&	3	&	72.2	$\pm$	2.7	\\
\hline
17  & 258.02078 & -38.51440	&	12.8	$\pm$	0.2	&	122	$\pm$	17	&	51	$\pm$	13	&	(4.2$\pm$0.6)	$\times$10$^{5}$	&	0.4	&	0	&	0	\\	
23  & 258.02648 & -38.51204	&	11.9	$\pm$	0.4	&	130	$\pm$	21	&	37	$\pm$	13	&	(4.5$\pm$0.7)	$\times$10$^{5}$	&	0.4	&	0	&	0	\\	
127	&	258.03146	&	-38.54054	&	10.8	$\pm$	0.3	&	80	$\pm$	17	&	12	$\pm$	5	&	(2.8$\pm$0.6)	$\times$10$^{5}$	&	0.4	&	0	&	0	\\		
 \hline 
\end{tabular}
\caption{Properties of the \herschel cores using the \getsources\,(+\getimages)\, algorithm. (1) Identification number,  (2,3) J2000 coordinates, (4) Envelope temperature, (5) Envelope mass, (6) Bolometric luminosity, (7) Volume density, (8) Jeans' mass, (9) Number of fragments inside the core, (10) Total mass of the fragments}    
\label{table:prop_cores}  
\tablefoot{
\tablefoottext{a}{The volume density was computed assuming a size of 0.1~pc and n$_{H_2}$=M$_{\rm{core}}$/($\mu m_H\times (4/3)\pi \times 0.1\rm{pc}^3 $).}
\tablefoottext{b}{SED fitting done without 160 and 250~$\mu$m, consequently, no aperture correction was done.}
}
\end{table*}
   
\section{Results and analysis}\label{sect:results}

The \herschel\,cores of RCW~120 \citep{fig17} were extracted with a new method using the \getimages\,\citep{men17} and \getsources\,algorithms (see Appendix~\ref{appendix:getsources} for details). The properties of the \herschel\,cores were obtained using a Spectral Energy Distribution (SED) fitting from 100 to 500~$\mu$m with the 160 or 250~$\mu$m flux mandatory to allow the aperture correction and better constrain the peak of the SED. The sources observed on the 3~mm ALMA image (Fig.~\ref{fig:rcw120_alma} bottom) were extracted with the same algorithms and we found 18 ALMA sources, called fragments, in total. To quantify the loss of emission due to the spatial filtering of the interferometer, we perform a SED fitting from 100 to 500~$\mu$m (\herschel data). The integrated intensity expected at 3~mm in the primary beam area is 0.3~Jy while the integrated intensity obtained from the observation is 0.06~Jy. Thus, we recover 20\% of the emission with the ALMA 12~m observations while the emission above 14\arcsec\, is filtered and lost. Using \herschel data, \citet{fig17} found that the \herschel cores belonging to the Condensation 1 of RCW~120 are the youngest and most massive of the region with in average of 124~M$_{\odot}$. Table~\ref{table:prop_cores} lists the properties of the \herschel\,cores where one ALMA fragment at least is detected. At a spatial resolution of 0.01~pc, the most massive core (Core 2, 376~M$_{\odot}$) is also the most fragmented with 5 sources found inside (see Fig.~\ref{fig:rcw120_fragments}). No fragments are observed towards Core 17, 23 and 127 which could be explained by the dilution above the maximum recoverable spatial scale of the interferometer if the cored have not yet undergone gravitational collapse. A zoom on the 18 fragments inside the 8 cores are presented in Fig.~\ref{fig:rcw120_fragments}. Assuming optically thin emission, the mass of the fragments was calculated using the Hildebrand formula \citep{hil83}: $M_{frag}=S_{3mm}\times R\times D^2/\kappa _{3mm}\times B_{3mm}(T_{dust})\times \Omega_{beam}$ where $S_{3mm}$ is the integrated flux at 3~mm (in Jy), $R$ is the gas-to-dust ratio assumed to be equal to 100, $D$ is the distance to the region (in pc), $B_{3mm}(T_{dust})$ is the Planck function at 3~mm (in Jy~sr$^{-1}$) assuming a temperature $T_{dust}$ (in K), $\Omega_{beam}$ is the beam solid angle of the observation (in sr) and $\kappa _{3mm}$ is the opacity at 3~mm assuming that $\kappa _{3mm}$=$\kappa _{300\mu m}\times (300\mu m/3mm)^{2}$ and $\kappa _{300\mu m}=10~\rm{cm}^2~g^{-1}$ \citep{hil83}. Different opacity values can be taken depending on the model used and the physical conditions of the medium \citep{oss94,pre93}. The choice of the opacity factor leads to a factor of 2 for the absolute mass uncertainty of the ALMA fragments \citep{mar12,deh12,cse17}. The temperature of the ALMA fragments is not well-defined and most of the studies consider a temperature around $\sim$20-25~K \citep{pal14,rat15,cse17,gin18}. This temperature can be justified if we consider that star formation already began in the fragments but since we do not know yet their evolutionary stage, we will compute the mass of the fragments assuming that they are at the same temperature than their hosting \herschel core (Louvet et al. submitted). The size of the fragments, computed as the geometrical mean of the major and minor angular size of the source, was deconvolved in quadrature with the beam angular resolution. The physical parameters of the ALMA sources are listed in Tab.~\ref{table:prop_fragments}.

%\begin{equation}\label{eq:hild}
%M_{frag}=\frac{S_{3mm}\times R\times D^2}{\kappa _{3mm}\times B_{3mm}(T_{dust})\times \Omega_{beam}}
%\end{equation}

\section{Discussion}

\subsection{Fragmentation inside the cores and thermal Jeans mass}\label{subsect:fragmentation}

In order to study the high-mass star formation process at the edges of \HII\,regions, we use these high resolution observations to characterize the fragmentation of the cores. When a core reaches the Jeans mass (M$_{\rm{Jeans}}$), it becomes gravitationally unstable. We expect the mass of the fragments to be of the order of M$_{\rm{Jeans}}$ after the fragmentation process. This mass limit depends on the sound speed and the density of the core which can be translated into temperature and density \citep{jea02,kip12}: $\rm{M}_{\rm{Jeans}}=0.6285\times (T_{\rm{env}}/{10K})^{1.5}\times(n_{H_2}/10^5cm^{-3})^{-0.5}$. The volume density of the \herschel\,cores was computed using the formula indicated in the footnote of Tab~\ref{table:prop_cores}. Since the size of several \herschel\,cores cannot be deconvolved, we assumed a typical scale of 0.1~pc which gives an upper limit for M$_{\rm{Jeans}}$. The resulting M$_{\rm{Jeans}}$ for our \herschel\,cores are of the same order of magnitude compared to the massive cores studied in \citet{pal14} or Louvet et al. (submitted) and range from 0.3~$M_{\odot}$ to 2.5~$M_{\odot}$ with an average of 0.8~$M_{\odot}$ (see Tab.~\ref{table:prop_cores}). From the mass of the cores and their corresponding M$_{\rm{Jeans}}$, we would have expected the fragmentation seen at 0.01~pc resolution to be more important. In particular, three of the {\textit{Herschel}} cores (17, 23 and 127) do not present any sign of fragmentation despite their associated low M$_{\rm{Jeans}}$. Fragmentation at this spatial scale, referred as multiplicity has been observed by \citep{lee15,pal18} in L1448N and OMC-1S for instance. Moreover, it is known that high-mass stars are often formed as part of a binary system \citep{zin07}. Therefore, it is possible that our sample of ALMA sources are sub-fragmented at a scale below 0.01~pc. The additional increase of the cores' temperature due to the heating provided by the proximity to the \HII\,region (if any) appears to be insignificant in increasing the Jeans mass threshold. Most of the theoretical models related to the fragmentation mechanisms consider other ingredients which are known to effectively prevent the fragmentation such as the turbulence \citep{pad01,fed10} and the magnetic field \citep{gir13,fon16,fon18}. In RCW~120, it has been shown that the compression of the \HII\,region has an effect on the millimetric-wave condensation studied in this work through a column-density PDF analysis \citep{tre14} and could have induced the gravitational collapse of the cores. We computed the  expected Jeans mass accounting for the turbulence (M$_{\rm{Jeans}}^{\rm{turb}}$) using the model of \citet{mac04} where the compression resulting from supersonic turbulence increases the volume density of the cores by a factor equal to the square of the Mach number. The thermal velocity dispersion is computed assuming a temperature equal to the average temperature of the cores. The non thermal velocity dispersion is then obtained by subtracting the thermal velocity dispersion from the N$_2$H$^+$(J=1$\rightarrow$0) velocity dispersion in quadrature (observed with the MALT90 survey at a spatial resolution of 30\arcsec-0.18~pc, \citealt{jac13})). Assuming a Mach number of 4 in the PDR of RCW~120 \citep{tre12}, we obtain an upper limit of 100~$M_{\odot}$ for the M$_{\rm{Jeans}}^{\rm{turb}}$ high enough to explain the mass of the fragments (Tab.~\ref{table:prop_fragments}).

\begin{table}    
\centering          
\begin{tabular}{|c|cccrr|}     % 7 columns 
\hline
\herschel & Id & $\alpha$ & $\delta$ & M$_{\rm{frag}}$ & Size  \\
 Core  & & \multicolumn{2}{c}{J2000 ($^{\circ}$)} & (M$_{\odot}$) & (mpc)\\
\hline
Core 1 & 8	&	258.0457	&	-38.5333	&	7.7	$\pm$	0.4	&	8.7	\\
& 12&	258.0440	&	-38.5321	&	2.8	$\pm$	0.3	&	3.1	\\
\hline
Core 2 &1	&	258.0363	&	-38.5129	&	27.4	$\pm$	0.6	&	8.6	\\
&5	&	258.0366	&	-38.5150	&	5.8	$\pm$	0.6	&	2.5	\\
&6	&	258.0358	&	-38.5124	&	19.1	$\pm$	1	&	8.8	\\
&10&	258.0356	&	-38.5163	&	5.1	$\pm$	0.7	&	4.3	\\
&11	&	258.0354	&	-38.5138	&	15.7	$\pm$	0.6	&	10.8	\\
\hline
Core 9 & 2	&	258.0369	&	-38.5264	&	8.2	$\pm$	0.7	&	1.6	\\
& 3	&	258.0370	&	-38.5269	&	17.6	$\pm$	0.9	&	9.3	\\
\hline
Core 10 & 16&	258.0471	&	-38.5204	&	7.7	$\pm$	0.8	&	9.6	\\
&17&	258.0450	&	-38.5194	&	7.8	$\pm$	0.6	&	9.9	\\
\hline
Core 11 & 7	&	258.0404	&	-38.5391	&	7.4	$\pm$	0.5	&	6.9	\\
\hline
Core 13 & 4	&	258.0334	&	-38.5386	&	4.8	$\pm$	0.4	&	3.1	\\
& 14&	258.0337	&	-38.5401	&	2.1	$\pm$	0.5	&	10.0\tablefootmark{a}	\\
\hline
Core 15 & 15&	258.0403	&	-38.5210	&	3.5	$\pm$	0.6	&	6.0	\\
\hline
Core 39 & 9	&	258.0450	&	-38.5248	&	8.7	$\pm$	0.4	&	8.1	\\
&13&	258.0448	&	-38.5265	&	31.6	$\pm$	0.7	&	22.4	\\
&18&	258.0457	&	-38.5236	&	31.8	$\pm$	1.5	&	28.8	\\
\hline
\end{tabular}
\caption{Properties of the fragments detected with ALMA at 3~mm. (1) Fragment's Id, (2,3) J2000-coordinates, (4) Mass assuming the temperature of the hosting core (see Table~\ref{table:prop_cores}), (5) Deconvolved size}   
\label{table:prop_fragments}  
\tablefoot{
\tablefoottext{a}{The size of this source cannot be deconvolved}
}
\end{table}

\subsection{High-mass star formation towards RCW~120}\label{subsect:massivestars_rcw120}

The millimetric-wave condensation studied in this work was the most promising place to search for the next generation of high-mass stars in RCW~120 since most of the massive cores observed with \herschel are located there. To know if high-mass star formation is likely to occur in the sample of \herschel cores, we use the criterion of \citet{bal17} where high-mass star formation can begin only if M$_{\rm{core}}$>1282$\left(r/[pc]\right)^{1.42}$~$M_{\odot}$. With a characteristic size of 0.1~pc, the mass limit at which a core can give rise to high-mass star is 50~$M_{\odot}$. In the sample of \herschel cores, 7 of them have a mass higher than this limit, accounting from the mass uncertainty of a factor of 2.Therefore, Cores 1, 2, 9, 10, 11, 15 and 39 host fragments which can potentially become high-mass stars depending on the star formation efficiency. Cores 17, 23 and 127 could form high-mass stars but no fragments are detected towards them, which means that the gravitational collapse has not occured yet. In Fig.~\ref{fig:rcw120_ch3so2}, we show two other molecular lines observed at the same spatial resolution with a detection set at 3$\sigma$ towards the most massive \herschel\,core. CH$_3$CN is a tracer of hot cores during the early-stage of star formation \citep{liu15}; SO$_2$ is a good tracer of outflows \citep{wri96,min16} and is also associated with Ultra Compact (UC) \HII\,regions \citep{min12,zha14,gal09}. The presence of CH$_3$CN and SO$_2$ emission associated with the high-mass Fragment 1 ($\sim$28~$M_{\odot}$) might indicate the presence of an UC\HII\,region in the PDR of RCW~120. This phenomenom can be compared to the RCW~79 \HII\,region where a C\HII\,region is clearly seen in the MSX observation \citep{zav06} towards the PDR. Most of the other fragments do not show any CH$_3$CN/SO$2$ lines emission which is the sign that the protostars are still forming and are in a earlier evolutionnary stage compared to the Fragment 1. Some of them have a mass well above the 8~$M_{\odot}$ limit and can potentially form high-mass stars whereas the low-mass fragments showing no signs of stellar activity for the moment, could accrete more mass from the molecular material contained in the core \citep{oha16} allowing them to exceed the 8~$M_{\odot}$ limit or will form low-mass stars. The fragmentation occuring at a spatial resolution of 0.01~pc reveals a distribution of sources with a mass ranging from 2 to 32~$M_{\odot}$ and an average of 12~$M_{\odot}$ in RCW~120. Compared to $\rho$-Ophiuchi where no \HII\,feedback is observed (\citealt{mot98} with a resolution of 0.008~pc) the mass of the fragments in RCW~120 is around one order of magnitude higher, which could indicate that the \HII\,region might favor the formation of high-mass stars. Nonetheless, recent ALMA follow-up of massive \herschel\,cores towards the high-mass star forming region NGC~6334, where several \HII\,regions are observed, was performed with a spatial resolution of 5~mpc (half of the spatial resolution of our observations) and a sensitivity of 0.11~$M_{\odot}$ by Louvet et al. (submitted). While the size of the NGC~6334 fragments, extracted with \getsources\,are comparable with the ALMA fragments in this study, the range of mass is completely different, from 0.2 to 2.6~$M_{\odot}$, with an average of 0.9$\pm$0.03~$M_{\odot}$ for NGC~6334. Using the same spectral index as in their study ($\beta$=1.5), the RCW~120 fragments' mass decreases but is still higher. In particular, the mass of the fragment 1 in Core 2 with this new spectral index is equal to 8~$M_{\odot}$, which is still in agreement with a possible high-mass star associated with an UC\HII\,region. Massive cores in this region have the potential to form high-mass stars and are subject to the \HII\,region feedback but contrary to RCW~120, no fragments are massive enough to form high-mass stars at this present moment. These two cases with opposite results indicate more observations are needed to understand the exact role of \HII\,regions in the formation of high-mass stars. Other observations towards the TUKH122 prestellar core at 3~mm and OMC-1S at 1~mm and 0.01~pc resolution \citep{oha18,pal18} where known \HII\,regions are present \citep{bal11} show an agreement with the thermal Jeans mechanism and a distribution of mass one order of magnitude lower compared to RCW~120.
 
\section{Conclusions}\label{sect:conclusions}

Using the 12~m ALMA antennas, we observed the densest millimetric-wave condensation of RCW~120 where most of the massive cores are found. We studied the properties of the 18 ALMA fragments found inside the cores detected with {\it{Herschel}} whose mass goes up to 32~$M_{\odot}$ with a detection limit of 0.12~$M_{\odot}$. The cores host from 0 to 5 fragments, the most massive one (Core 2) being the most fragmented. The high fragmentation expected from the Jeans instabilities is not observed towards these massive cores and is in favor of the formation of massive stars and the addition of turbulence or magnetic field. CH$_3$CN and SO$_2$ emission are observed towards the main fragment in the most massive core arguing for the presence of an UC\HII\,region in the PDR. Therefore, these new ALMA observations have shown that Condensation 1 of RCW~120 is a favourable place for the formation of high-mass stars. 

\begin{acknowledgements}
LB acknowledges support from CONICYT Project PFB-06. This paper makes use of the following ALMA data: ADS/JAO.ALMA\#2016.1.00314.S. ALMA is a partnership of ESO (representing its member states), NSF (USA) and NINS (Japan), together with NRC (Canada), MOST and ASIAA (Taiwan), and KASI (Republic of Korea), in cooperation with the Republic of Chile. The Joint ALMA Observatory is operated by ESO, AUI/NRAO and NAOJ.
\end{acknowledgements}

   \bibliographystyle{aa} % style aa.bst
   \bibliography{biblio} % your references Yourfile.bib

\clearpage
\onecolumn
\begin{appendix}

\section{Images of ALMA fragments for each cores}

%\vspace*{3cm}

\begin{figure}
\centering
\includegraphics[width=0.8\linewidth]{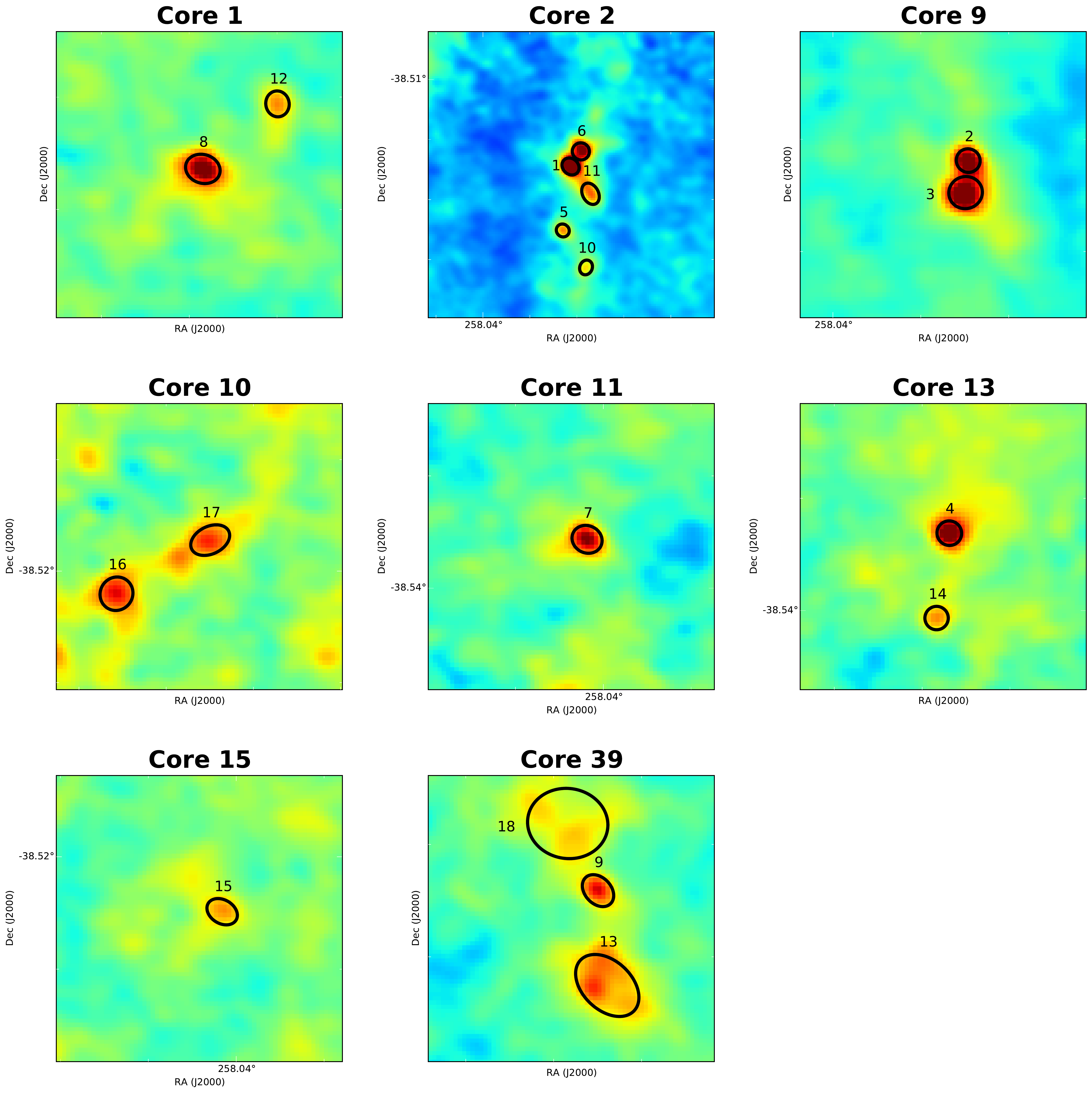}
\caption{ALMA sources extracted with \getsources\,(black ellipses) for each of the \herschel cores at 3~mm.}
\label{fig:rcw120_fragments}
\end{figure}

\begin{figure}
\centering
\includegraphics[width=0.6\linewidth]{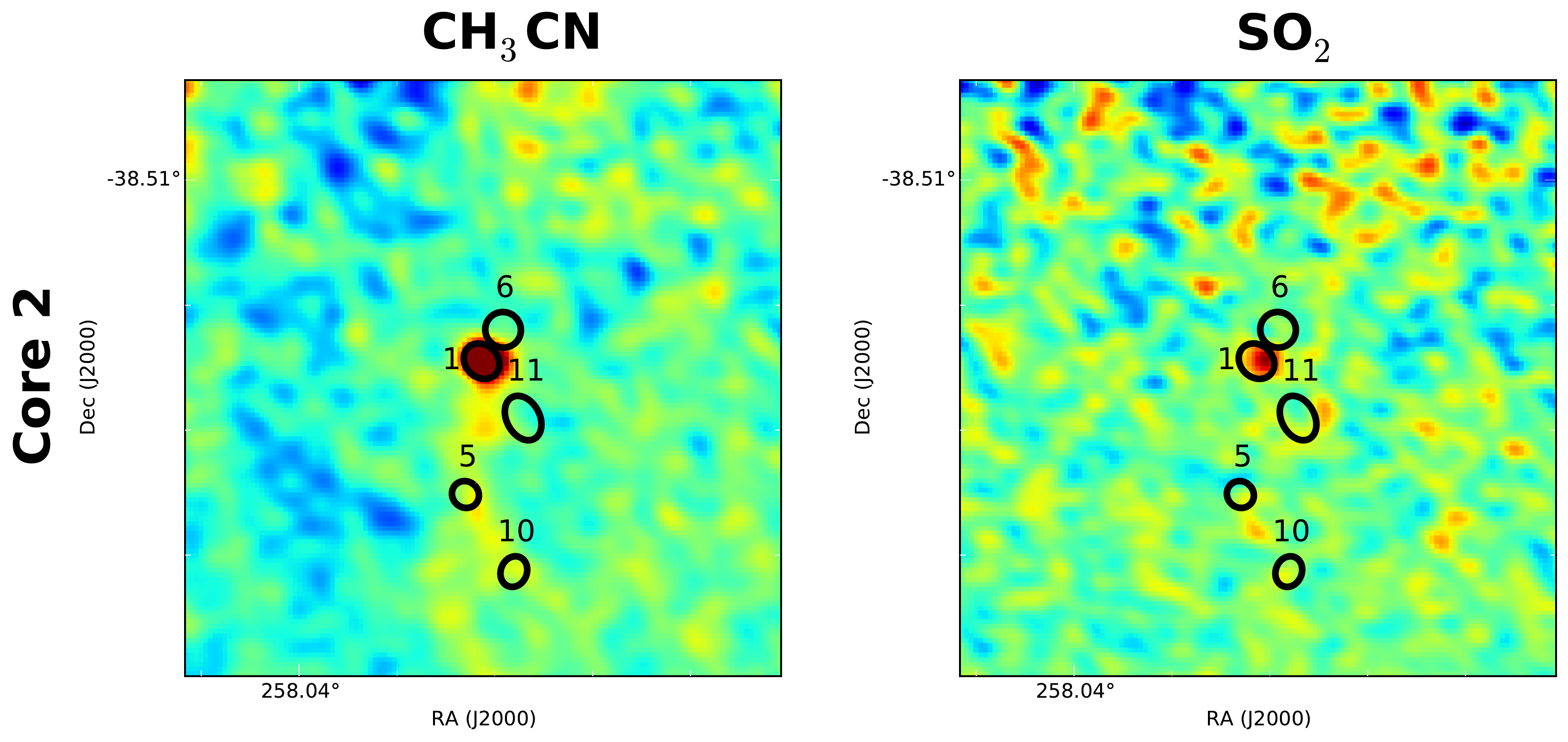}
\caption{Integrated emission of the CH$_3$CN and SO$_2$ molcular lines towards the \herschel\,Core 2}
\label{fig:rcw120_ch3so2}
\end{figure}

\twocolumn

\section{\getsources\,extraction}\label{appendix:getsources}

The \getsources\,algorithm is a multi-wavelength multi-scale extraction program used to extract compact sources and filaments through a single-scale decomposition approach. The original images were processed using the prepareobs script which creates a set of images used for the detection and the measurements steps. Firstly, \getsources\,decomposes the images at each wavelength into a set of single spatial scales. Sources whose size does not correspond to the spatial scale are filtered while the visibility of the others is enhanced. This sample of images is then cleaned using an intensity thresholding method. Once the cleaning is done, each set of single spatial scale images are combined to form independent wavelength images. In practice, two sets of combined images are created, one to follow the evolution of the source's shape (footprint) and the other one to follow the evolution of the peak intensity. Finally, the algorithm looks for every set of 4-connected pixels shape and assigns a source identification number to each of them. The second main step of the process consists of measuring the properties of each detected source and at each wavelength. In order to improve the sources' detection and measurements, a new set of flat detection images are created using the results from the first-pass in \getsources. The entire steps of detections and measurements mentionned above are performed again using these flat images as input detection images.\\
Recently, extractions made with the \getsources\,algorithm were improved by the use of \getimages\,which replaces the first-pass in \getsources. Using a median filtering on different spatial scales ranging from the observational beam to the maximal size of the structures that we want to detect, \getimages\,provides a background-subtracted flattened image where the small-scale fluctuations outside the structures of interest are uniform. Based on that image and on the maximal size of the structures one wants to extract, the \getsources\,algorithm is used to obtain the catalog of sources together with their properties.
The \herschel cores from \citet{fig17} were extracted using the \getsources\,algorithm and the HOBYS recipe \citep{tig17}. Since the use of \getimages\,improves the quality of the final detections catalog compared to the previous extractions, the cores were extracted again following the new strategy by combining the \getsources\,and \getimages\,algorithms as well as the ALMA fragments. For this work, we used \getimages\,(Version 2.171128) with the default parameters and a maximal size of the structures set to 2\arcsec\,, and \getsources\,(Version 1.140127) with the default parameters. The configuration files of \getimages\, and \getsources\, can be found together with the catalog on the CDS and more explanations about the parameters can be found in \citet{men12,men17}. While 13 cores are found in the former extraction (\getsources\,only), only 10 cores are listed in the new \herschel catalog (\getsources\, \& \getimages). The Core 58 is not detected in the new extraction and the Core 15 and 37 are detected but not selected due to the lack of reference flux for the aperture correction which is mandatory in the HOBYS recipe. Nonetheless, since an ALMA source is found for the Core 15, we keep it for our analysis. However, one has to remember that no aperture correction was made for the SED of this source, which could lead to higher uncertainties for its derived physical properties.

\end{appendix}

\end{document}